\documentclass[aps,prl,twocolumn,superscriptaddress,showpacs,preprintnumbers,amsmath,amssymb]{revtex4}
\usepackage{graphicx} 
\usepackage{dcolumn}  
\usepackage{epsfig}
\usepackage{color}
\usepackage{ulem}

\graphicspath{{ps}}

\begin{document}
\preprint{\vbox{ \hbox{   }
}}
\title
{\quad\\[0.5cm] Observation of a new $D_{sJ}$ meson in
$B^{+}  \! \to  \! \bar{D}^{0} D^{0} K^{+}$ decays}
\affiliation{Budker Institute of Nuclear Physics, Novosibirsk}
\affiliation{Chiba University, Chiba}
\affiliation{University of Cincinnati, Cincinnati, Ohio 45221}
\affiliation{Department of Physics, Fu Jen Catholic University, Taipei}
\affiliation{Justus-Liebig-Universit\"at Gie\ss{}en, Gie\ss{}en}
\affiliation{The Graduate University for Advanced Studies, Hayama}
\affiliation{Gyeongsang National University, Chinju}
\affiliation{Hanyang University, Seoul}
\affiliation{University of Hawaii, Honolulu, Hawaii 96822}
\affiliation{High Energy Accelerator Research Organization (KEK), Tsukuba}
\affiliation{Hiroshima Institute of Technology, Hiroshima}
\affiliation{University of Illinois at Urbana-Champaign, Urbana, Illinois 61801}
\affiliation{Institute of High Energy Physics, Chinese Academy of Sciences, Beijing}
\affiliation{Institute of High Energy Physics, Vienna}
\affiliation{Institute of High Energy Physics, Protvino}
\affiliation{Institute for Theoretical and Experimental Physics, Moscow}
\affiliation{J. Stefan Institute, Ljubljana}
\affiliation{Kanagawa University, Yokohama}
\affiliation{Korea University, Seoul}
\affiliation{Kyungpook National University, Taegu}
\affiliation{\'Ecole Polytechnique F\'ed\'erale de Lausanne (EPFL), Lausanne}
\affiliation{University of Ljubljana, Ljubljana}
\affiliation{University of Maribor, Maribor}
\affiliation{University of Melbourne, School of Physics, Victoria 3010}
\affiliation{Nagoya University, Nagoya}
\affiliation{Nara Women's University, Nara}
\affiliation{National Central University, Chung-li}
\affiliation{National United University, Miao Li}
\affiliation{Department of Physics, National Taiwan University, Taipei}
\affiliation{H. Niewodniczanski Institute of Nuclear Physics, Krakow}
\affiliation{Nippon Dental University, Niigata}
\affiliation{Niigata University, Niigata}
\affiliation{University of Nova Gorica, Nova Gorica}
\affiliation{Osaka City University, Osaka}
\affiliation{Osaka University, Osaka}
\affiliation{Panjab University, Chandigarh}
\affiliation{RIKEN BNL Research Center, Upton, New York 11973}
\affiliation{Saga University, Saga}
\affiliation{University of Science and Technology of China, Hefei}
\affiliation{Seoul National University, Seoul}
\affiliation{Sungkyunkwan University, Suwon}
\affiliation{University of Sydney, Sydney, New South Wales}
\affiliation{Tata Institute of Fundamental Research, Mumbai}
\affiliation{Toho University, Funabashi}
\affiliation{Tohoku Gakuin University, Tagajo}
\affiliation{Tohoku University, Sendai}
\affiliation{Department of Physics, University of Tokyo, Tokyo}
\affiliation{Tokyo Institute of Technology, Tokyo}
\affiliation{Tokyo Metropolitan University, Tokyo}
\affiliation{Tokyo University of Agriculture and Technology, Tokyo}
\affiliation{Virginia Polytechnic Institute and State University, Blacksburg, Virginia 24061}
\affiliation{Yonsei University, Seoul}
  \author{J.~Brodzicka}\affiliation{High Energy Accelerator Research Organization (KEK), Tsukuba} 
 \author{H.~Palka}\affiliation{H. Niewodniczanski Institute of Nuclear Physics, Krakow} 
  \author{I.~Adachi}\affiliation{High Energy Accelerator Research Organization (KEK), Tsukuba} 
  \author{H.~Aihara}\affiliation{Department of Physics, University of Tokyo, Tokyo} 
  \author{V.~Aulchenko}\affiliation{Budker Institute of Nuclear Physics, Novosibirsk} 
  \author{A.~M.~Bakich}\affiliation{University of Sydney, Sydney, New South Wales} 
  \author{E.~Barberio}\affiliation{University of Melbourne, School of Physics, Victoria 3010} 
  \author{A.~Bay}\affiliation{\'Ecole Polytechnique F\'ed\'erale de Lausanne (EPFL), Lausanne} 
  \author{I.~Bedny}\affiliation{Budker Institute of Nuclear Physics, Novosibirsk} 
  \author{U.~Bitenc}\affiliation{J. Stefan Institute, Ljubljana} 
% ...
\collaboration{The Belle Collaboration}

\begin{abstract}

We report the observation of a new $D_{sJ}$ meson produced in 
$B^{+}  \! \to \! \bar{D}^{0} D_{sJ}  \! \to \! \bar{D}^{0} D^{0} K^{+}$.
This state has a mass of $M=2708 \pm 9 ^{+11}_{-10}~\rm{MeV}/{\it c}^{2}$, 
a width $\Gamma = 108 \pm 23 ^{+36}_{-31} ~\rm{MeV}/ {\it c}^{2}$
and a $1^{-}$ spin-parity. 
The statistical significance of this observation is $8.4 \sigma$. 
The results are based on an analysis of $449$ million $B\bar{B}$ events 
collected at the $\Upsilon(4S)$ resonance
with the Belle detector at the KEKB asymmetric-energy $e^{+} e^{-}$ collider.

\end{abstract}

\pacs{14.40.Lb, 13.25.Hw, 13.20.Fc}

\maketitle

\tighten

{\renewcommand{\thefootnote}{\fnsymbol{footnote}}}
\setcounter{footnote}{0}

At the level of quark diagrams, the decay $B \! \to \! \bar{D}D K$ 
proceeds dominantly via the CKM-favored 
$\bar{b} \!  \to \!  \bar{c}  W^{+} \!  \to \! \bar{c}c\bar{s}$ 
transition. The transition amplitudes can be 
categorized as due to either external $W$-
or internal (color-suppressed) $W$-emission diagrams.
The decay $B^{+} \! \to \!  \bar{D}^{0}D^{0} K^{+}$~\cite{cc}
can proceed through both types of diagrams;
thus it is promising for searches for new
$c\bar{s}$ states as well as for some
$c\bar{c}$ states lying above $D^{0}\bar{D}^{0}$ 
threshold.   
The unexpected discoveries of the $D^*_{s0}(2317)$
and  $D_{s1}(2460)$ mesons show that our understanding of 
$c\bar{s}$ spectroscopy might be incomplete, while
experimental data on $c\bar{c}$ states with decay 
channels open to $D^{(*)}\bar{D}^{(*)}$ are scarce.

The decays $B  \! \to \! \bar{D}D K$ have been previously 
studied with a small data sample at LEP~\cite{aleph} and 
more recently a larger statistics exploratory study 
was performed by BaBar~\cite{babar}.
In this letter we report the first study of the
Dalitz plot of $B^{+} \! \to \! \bar{D}^{0} D^{0} K^{+}$. 
 
The study is performed using data 
collected  with the Belle detector at the KEKB asymmetric-energy
$e^+e^-$ (3.5 on 8~GeV) collider~\cite{KEKB},
operating at the $\Upsilon(4S)$ resonance ($\sqrt{s}=10.58$~GeV). 
The data sample corresponds to the integrated luminosity
of $414~{\rm fb}^{-1}$ and contains 
$449$ million $B\bar{B}$ pairs. 
The Belle detector is a large-solid-angle magnetic
spectrometer that is described in detail elsewhere~\cite{Belle}.

Well measured charged tracks are identified by combining information
from time-of-flight, Cherenkov and ionisation detectors. 
Requirements on the particle identification variable are imposed
that identify a charged kaon with $90\%$ efficiency, a charged pion with
almost $100\%$ efficiency and have less than 
$10\%$ $K \! \leftrightarrow \! \pi$ 
misidentification probability. 
Any track that is positively identified as an electron is rejected.  

Candidate $K^{0}_{S} \! \to \! \pi^{+}\pi^{-}$ decays are identified
by a displaced secondary vertex, a two-pion momentum vector that is
consistent with a $K^{0}_{S}$ originating from the IP and a $\pi^{+}\pi^{-}$ 
invariant mass within $\pm 15~\rm{MeV/{\it c}^{2}}~(\pm 3 \sigma)$ 
of the nominal $K^{0}_{S}$ mass. 
Candidate $\pi^{0}$ mesons are reconstructed from pairs of
identified photons, each with a minimum energy 
of $50~\rm{MeV}$, that have an invariant mass within 
$\pm 15~\rm{MeV/{\it c}^{2}}~(\pm 3 \sigma)$ of the $\pi^{0}$ mass.

$D^{0}$ mesons are reconstructed in the
$K^{-}\pi^{+}$, $K^{-}\pi^{+}\pi^{+}\pi^{-}$, $K^{-}\pi^{+}\pi^{0}$,
$K^{0}_{S}\pi^{+}\pi^{-}$ and $K^{-}K^{+}$ decay modes.
We preselect $D^{0}$ candidates using a
signal window $\pm 30~\rm{MeV/{\it c}^{2}}~(\pm 5 \sigma)$ 
around the nominal $D^{0}$ meson mass for all decay modes except for
$D^{0} \!\to \! K^{-}\pi^{+}\pi^{0}$, where 
a $\pm 50~\rm{MeV/{\it c}^{2}}~(\pm 5 \sigma)$ signal window is used.
Mass- and vertex-constrained fits are applied to $D^{0}$ 
candidates to improve their momentum resolution.

To suppress the continuum background 
($e^{+}e^{-} \! \to \! q\bar{q}$, $q=u,d,s,c$)
we require the ratio of the second to the zeroth 
Fox-Wolfram moments~\cite{fox-wolfram} to be less than $0.3$.

We form $\bar{D}^{0} D^{0} K^{+}$ combinations using
$D^{0}$ candidates with momenta in the $\Upsilon(4S)$ rest frame (cms) 
kinematically allowed in $B^{+} \! \to \! \bar{D}^{0} D^{0} K^{+}$.
The momenta of the secondaries from a $B$ meson candidate decay are
refitted to a common vertex with an interaction point (IP) 
constraint that takes into
account the $B$ meson decay length.
The $B$ meson candidates are identified by their cms energy difference,
$\Delta E = \Sigma_{i}E_{i} - E_{\rm beam}$,
and their beam constrained mass,
$M_{\rm bc} = \sqrt{E^{2}_{\rm beam} - (\Sigma_{i}\vec{p}_i)^2}$, 
where $E_{\rm beam}=\sqrt{s}/2$
is the beam energy in the cms and $\vec{p}_i$ and $E_i$ are
the three-momenta and energies of the $B$ candidate's decay products.
We select $B$ candidates with
$M_{\rm bc}>5.2$~GeV$/{\it c}^2$ and $-0.4~{\rm GeV} < \Delta E <0.3$~GeV.
Exclusively reconstructed $B^{+} \! \to  \! \bar{D}^{0} D^{0} K^{+}$ signal 
events have an $M_{\rm bc}$ distribution that peaks at the nominal $B$ 
meson mass; the $\Delta E$ distribution peaks at zero. 

We employ a discriminator (likelihood ratio) 
based on the $D^{0}$ meson signal significance
to select the unique $B$ candidate in the event, defined as:
${\cal LR}(M_{D^{0}}) = \frac{S(M_{D^{0}})}{S(M_{D^{0}})+B(M_{D^{0}})}$,
where $S$ and $B$ are the signal and the background likelihoods 
that depend on the $D^0$ candidate's invariant mass $(M_{D^{0}})$.
This discriminator is determined from fits to the $M_{D^{0}}$ 
distributions for each $D^{0}$ decay mode separately, 
using a data sample enriched in $B^+  \! \to \! \bar{D}^{0} D^{0} K^+$ 
decays. In these fits $S$ and $B$ are parametrized respectively by 
a double gaussian and a linear functions.
For events with multiple
$B^+  \! \to  \! \bar{D}^{0} D^{0} K^+$ candidates, the product
${\cal LR}_B = {\cal LR}(M_{\bar{D}^{0}}) \times {\cal LR}(M_{D^{0}})$
is calculated and the candidate with the largest ${\cal LR}_B$
is accepted. 
The  ${\cal LR}_B$ discriminator is also used to suppress
combinatorial backgrounds to $B^{+}  \! \to \! \bar{D}^{0} D^{0} K^{+}$
and to enhance the signal purity. 
  Monte Carlo (MC) studies showed that this selection, which does not rely 
  on the $\Delta E$ and $M_{bc}$ values used in the $B$ signal definition, 
  does not introduce biases.
  The fraction of events with multiple candidates is $45\%$ and the average 
  candidate multiplicity is 2.2. The main background is due 
  to $B\bar{B}$ production, with multiple candidates originating from wrong 
  pairing of $D^0$'s or swapped kaons. 

The $\Delta E$ and $M_{\rm bc}$ distributions 
for the $B^{+} \! \to  \! \bar{D}^{0}D^{0} K^{+}$ 
decay candidates, selected with  ${\cal LR}_B > 0.01$
requirement, are shown in Fig.~{\ref{dembc},
where the $\Delta E$ distribution 
is shown 
for 
$\mid \!  M_{\rm bc} - m_{B} \!  \mid <  3\sigma_{M_{\rm bc}}$ 
($\sigma_{M_{\rm bc}} = 2.7~{\rm MeV}/{\it c}^{2}$, 
$m_{B}$ is the nominal $B$ meson mass) and the $M_{\rm bc}$ distribution 
is shown for $\mid \! \Delta E \! \mid < 3\sigma_{\Delta E }$
($\sigma_{\Delta E } = 6.6~{\rm MeV}$).
\begin{figure}[!h]      
\vspace{-0.1cm}
\begin{center}      
\includegraphics[width=0.22\textwidth,height=0.20\textwidth]{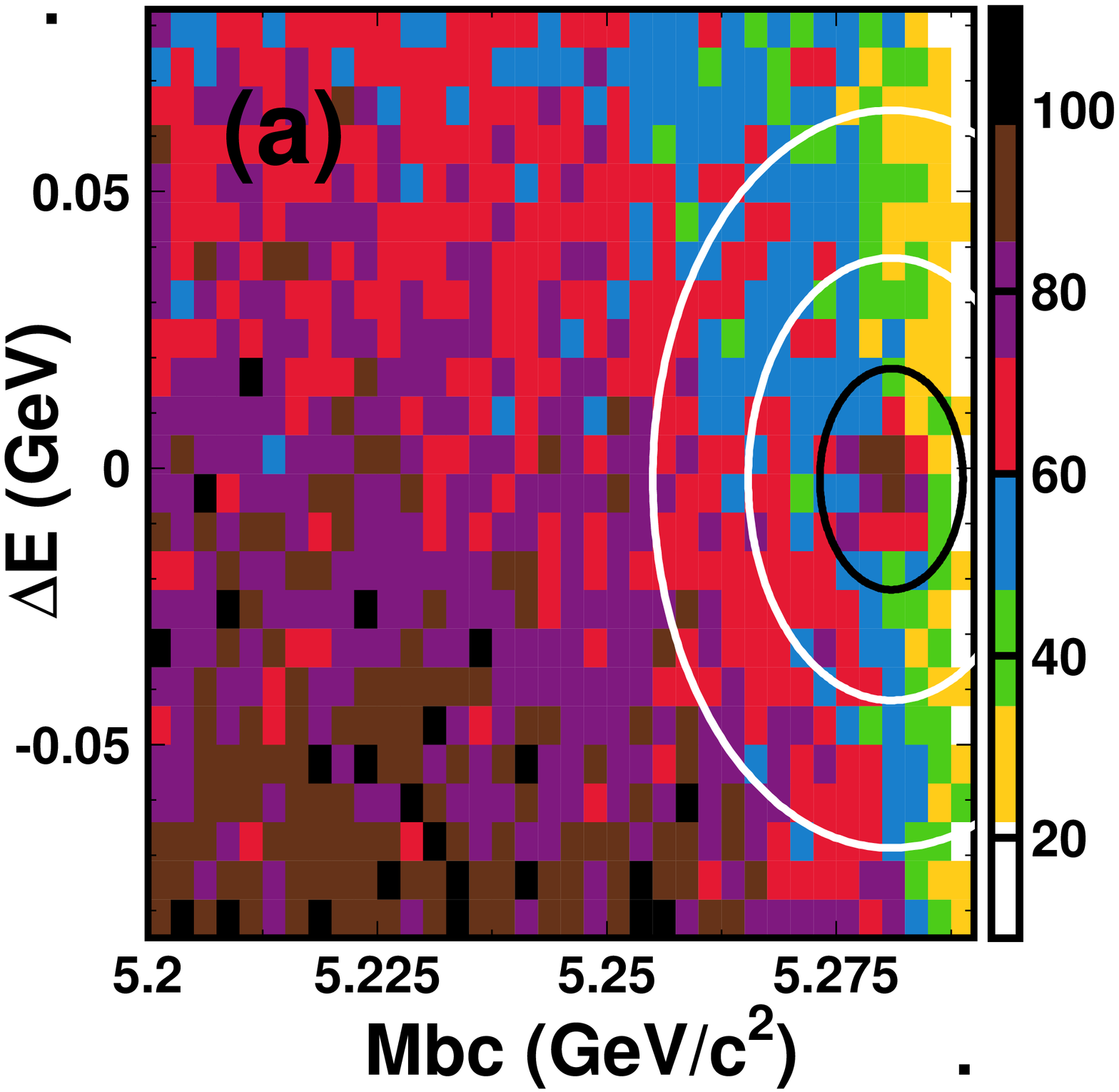}      
\includegraphics[width=0.21\textwidth,height=0.20\textwidth]{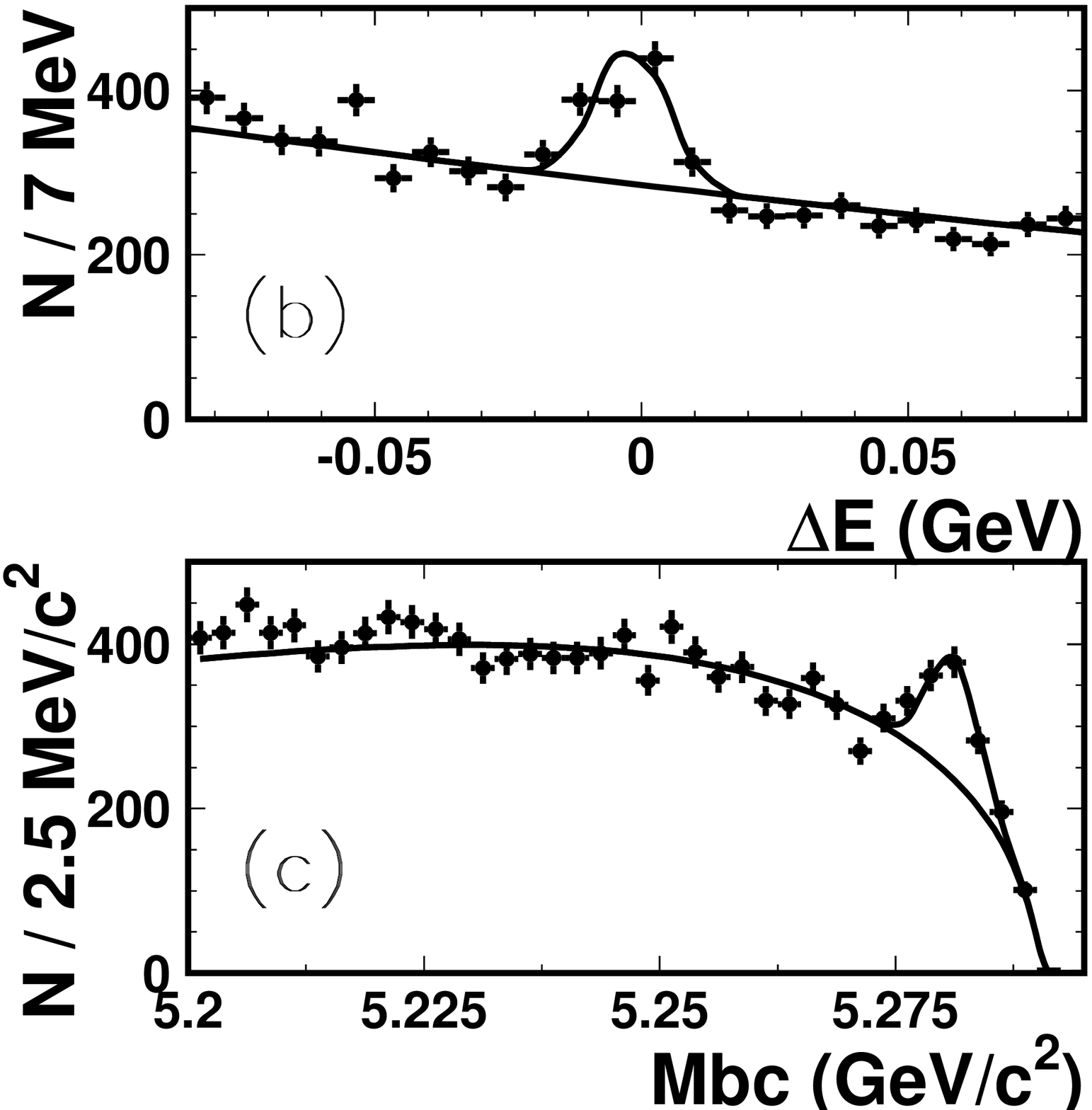}  
\caption{$\Delta E$~vs.~$M_{\rm bc}$ (a), $\Delta E$ (b) and $M_{\rm bc}$ (c) 
distributions for $B^{+} \! \to \! \bar{D}^{0} D^{0} K^{+}$. Black (white) ellipses 
in (a) enclose the signal (sideband) regions described in the text. }
\label{dembc}
\end{center}                                       
\vspace{-0.6cm}
\end{figure}      

From a study of the $M_{\rm bc}$ and $ \Delta E$ background distributions 
in large MC samples of generic $B\bar{B}$ and $q\bar{q}$ events
as well as $D^0$-mass sidebands in data, we find no significant
peaking background. 

To extract the signal yield,
we perform two-dimensional (2D) extended unbinned maximum-likelihood fits  
to $\Delta E$ and $M_{\rm bc}$.
The probability density functions (PDFs) for the $M_{\rm bc}$ and $\Delta E$ 
signals are Gaussians.
The background PDF for $M_{\rm bc}$ is represented by 
a phenomenological function~\cite{argus} with a phase-space-like
behaviour near the kinematic boundary; the
$\Delta E$ background is parameterized by a second-order polynomial.
The likelihood function is maximized with free parameters for the signal yield, 
the Gaussian means and widths, and four parameters that describe shapes 
of the background distributions. 
From the fit, we obtain a signal yield of $N_{\rm sig}= 399\pm 40$ 
events. The results of the fit are superimposed on
the $\Delta E$ and $M_{\rm bc}$ projections shown in 
Fig.~\ref{dembc}(b)-(c). 

We determine the branching fraction from the relation: 
${\cal B}(B^{+} \! \to \! \bar{D}^{0}D^{0}K^{+})\!=\!
\frac{N_{{\rm sig}}}{N_{B \bar{B}} \sum_{ij}\epsilon_{ij}
{\cal B}(\bar{D}^{0}\! \to \! i){\cal B}(D^{0}\! \to \! j)}$,
where $\epsilon_{ij}$ are efficiencies for the $D^{0}$ decay channels $i$ and $j$. 
$N_{B\bar{B}}$ is the number of analyzed $B\bar{B}$ pairs, 
$N_{B\bar{B}}= 449 \times 10^6$, 
and $N_{B^{+}B^{-}} = N_{B^{0}\bar{B}^{0}}$ is assumed.
The efficiencies are determined by MC using 
a model that reproduces the observed Dalitz plot features (discussed below).
The sum in the denominator of the above relation is 
$4.0\times 10^{-4}$.
We obtain
${\cal B}(B^{+} \! \to \! \bar{D}^{0}D^{0}K^{+})=
(22.2\pm 2.2^{+2.6}_{-2.4})\times 10^{-4}$,
where the first error is statistical and the second is
systematic.  The latter includes contributions due to 
uncertainties in the efficiency determination
(tracking and particle identification efficiency, data-MC differences in 
$\Delta E, M_{\rm bc}$ signal shapes), 
the ${\cal LR}_B$ selection, the background parameterization, 
the MC model used in the efficiency calculation, 
the intermediate $D^{0}$ branching fractions and $N_{B \bar{B}}$. 
This result supersedes our previous determination~\cite{chistov}, 
which assumed a phase space model in the efficiency determination.
\begin{figure*}[!ht]      
\vspace{-0.1cm}
\begin{center}      
\includegraphics[width=0.24\textwidth,height=0.19\textwidth]{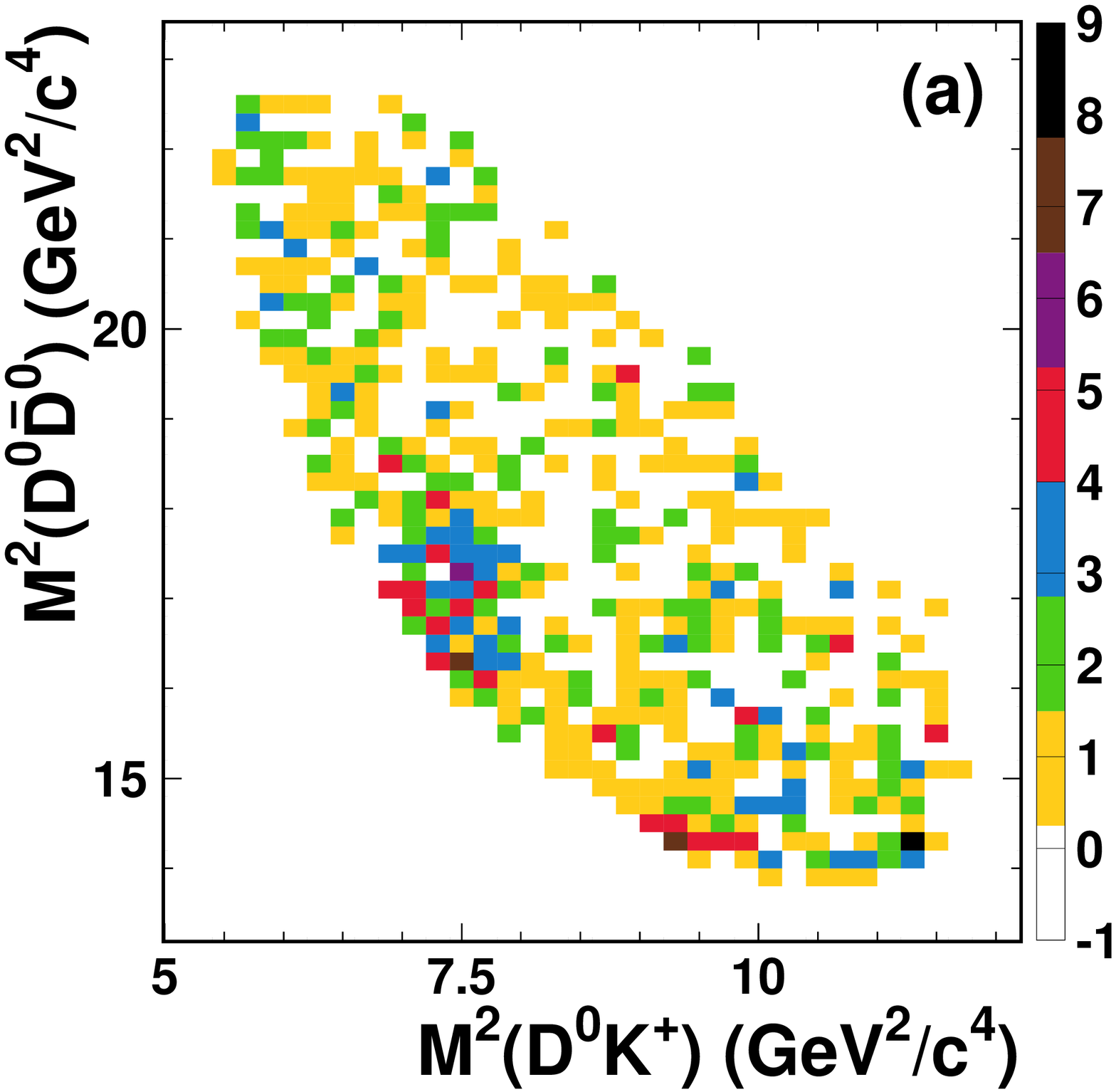}
\includegraphics[width=0.23\textwidth,height=0.19\textwidth]{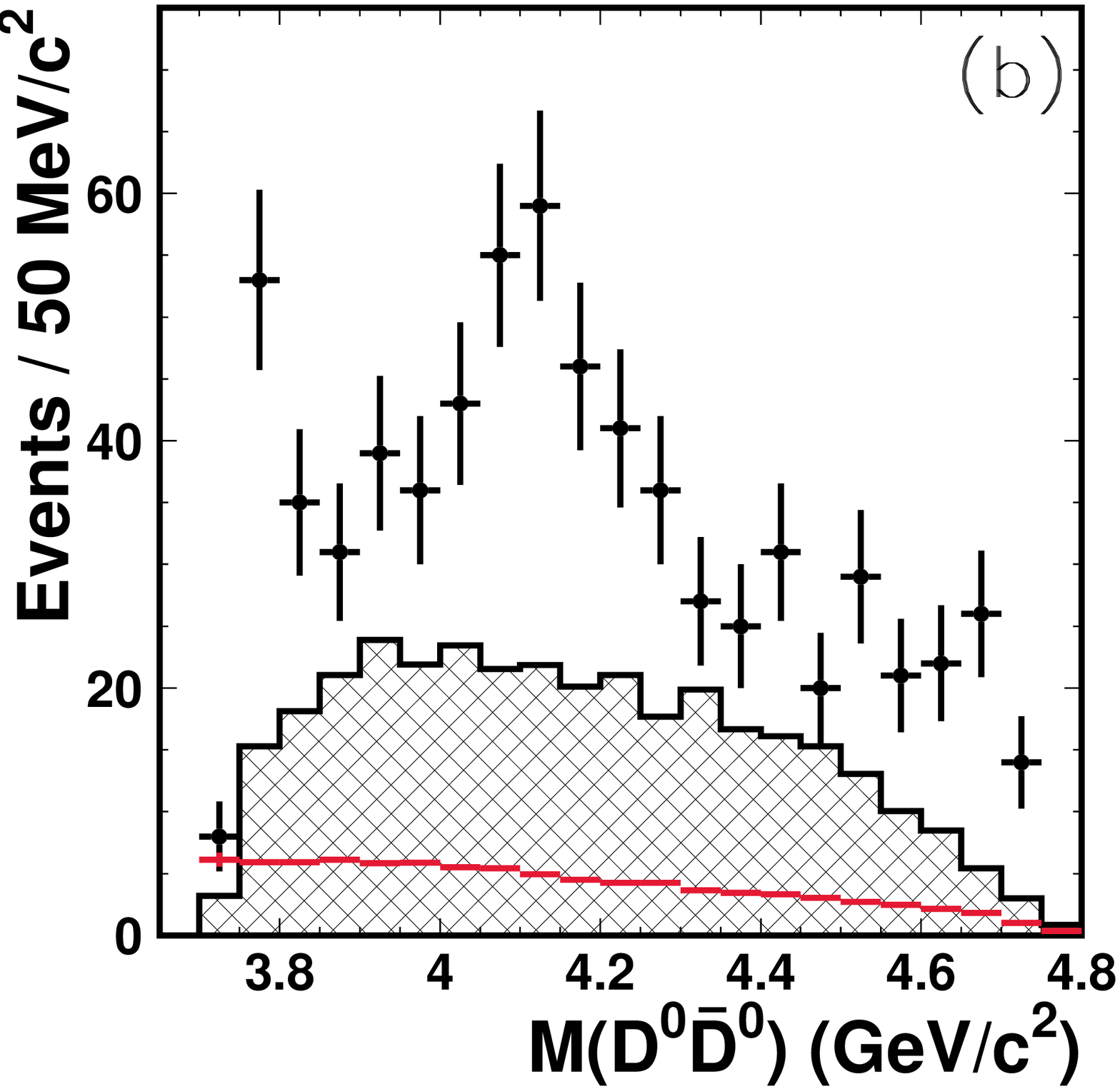}      
\includegraphics[width=0.21\textwidth,height=0.19\textwidth]{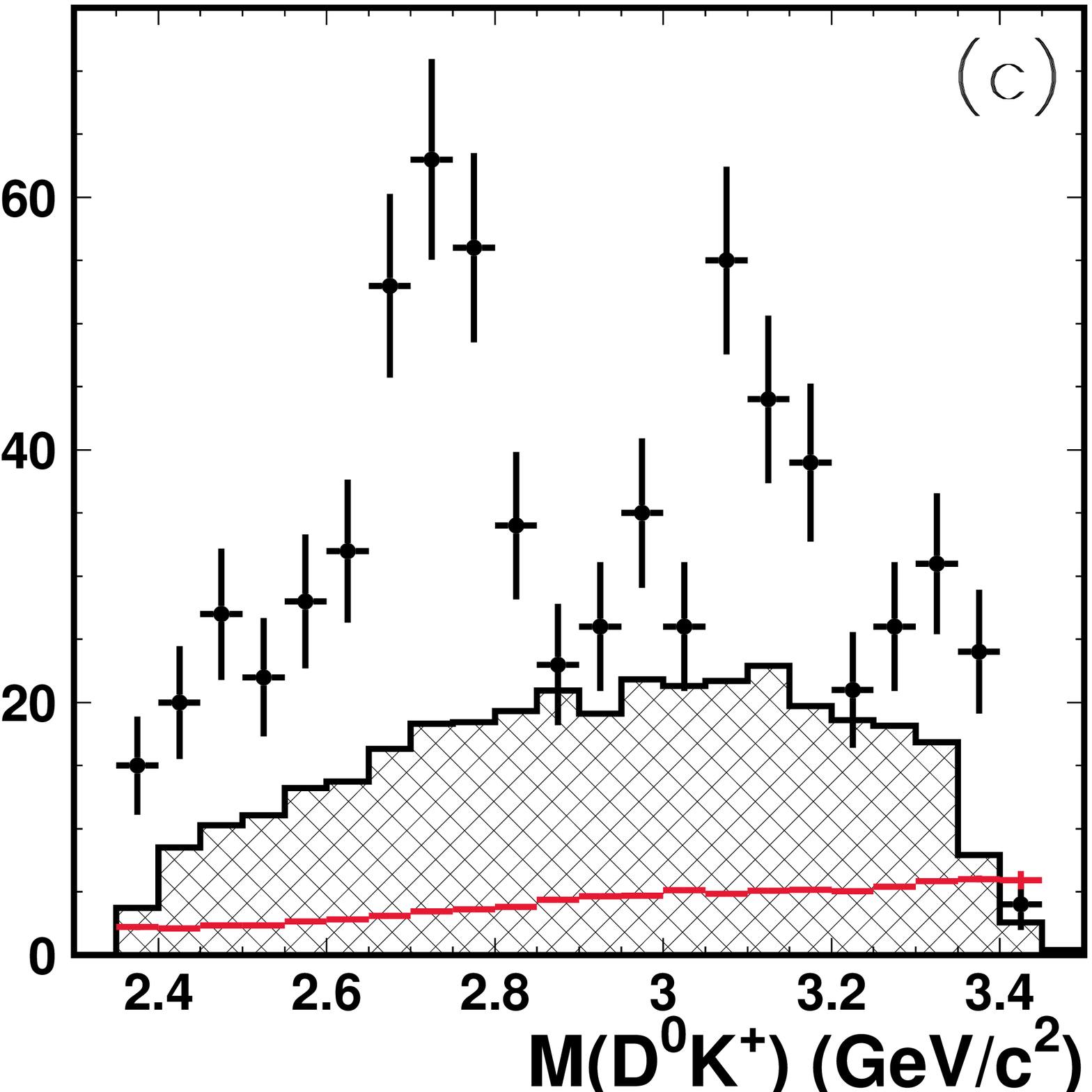}      
\includegraphics[width=0.233\textwidth,height=0.19\textwidth]{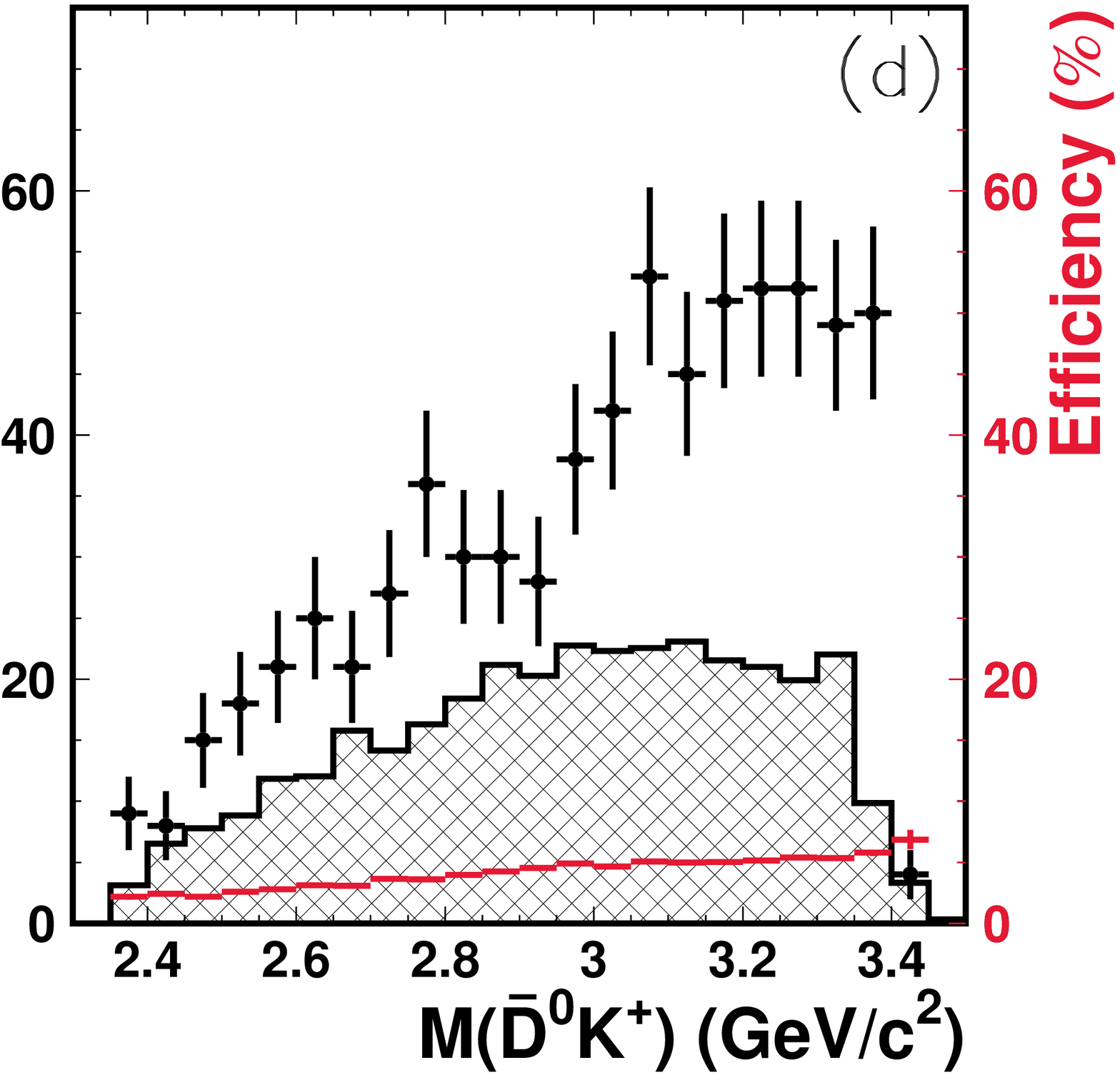}
\caption{Dalitz plot (a) and projections for 
$B^{+} \! \to \! \bar{D}^{0} D^{0} K^{+}$ 
in the $1.5\sigma ~\Delta E \!- \! M_{\rm bc}$ signal region:
           $M(D^{0} \bar{D}^{0})$ (b),
           $M(D^{0}K^{+})$ (c), 
           $M(\bar{D}^{0}K^{+})$ (d). 
Hatched histograms represent background, 
red/solid curves show the efficiency.}
\label{dalitz_d0d0bkc}
\end{center}                                       
\vspace{-0.6cm}
\end{figure*}   

The main features of the data can be seen in
the Dalitz plot 
$M^{2}(D^{0}\bar{D}^{0})$~vs.~$M^{2}(D^{0}K^{+})$
for events from a signal region defined by the ellipse
$R^2 \! \equiv \! (\Delta E /\sigma_{\Delta E})^{2}+
((M_{\rm bc}-m_{B})/\sigma_{M_{\rm bc}})^{2}$, $R^2 \! < \! (1.5) ^2$ 
shown in Fig.~\ref{dalitz_d0d0bkc}(a).
The three two-body invariant mass distributions are 
shown in Figs.~\ref{dalitz_d0d0bkc}(b)-(d).
The hatched histograms 
represent the background distributions obtained for events
from an elliptical strip surrounding the $\Delta E \!- \!M_{\rm bc}$
signal region, defined by $6^2 \! < \! R^2 \! < \! 10^2$.
The background distributions are normalized to the number
of background events under the signal 
peak $(\pm 1.5\sigma)$ as determined from the 2D
$\Delta E$ and $M_{\rm bc}$ fit.
The data are not efficiency corrected.
The efficiency as a function of invariant mass is shown in 
Figs.~\ref{dalitz_d0d0bkc}(b)-(d) as a continuous curve. 

A pronounced feature of the Dalitz plot 
is the accumulation of events in the region
$16~{\rm GeV}^{2}/{\it c}^{4}  \! \leq  \!
M^{2}(D^{0} \bar{D}^{0}) \! \leq  \! 18~{\rm GeV}^{2}/{\it c}^{4}$ 
and
$7~{\rm GeV}^{2}/{\it c}^{4}  \! \leq  \! M^{2}(D^{0}K^{+})  \! \leq  \! 
8~{\rm GeV}^{2}/{\it c}^{4} $
possibly due to the overlap of 
a horizontal band that could be due to the $\psi(4160)$, 
$\psi(4040)$ and a vertical band that
cannot be attributed to any known $c \bar{s}$ state. A horizontal band at 
$M^{2}(D^{0}\bar{D}^{0}) \simeq 14.2~{\rm GeV}^{2}/{\it c}^{4}$ 
corresponds to $\psi(3770)$ production.  

The distributions in Fig.2 are meant 
only to illustrate the features of the data.
In the subsequent analyses a more robust procedure was used to
obtain background-subtracted mass 
distributions.
We determine the $\Delta E$ vs. $M_{\rm bc}$ distributions
for events from mass bins of the Dalitz plot projection and fit 
the signal and background shapes to obtain 
$B$ meson signal yield vs. invariant mass.
$20~\rm{MeV}/{\it c}^{2}$ mass bins are used for 
the $\psi(3770)$, while $50~\rm{MeV}/{\it c}^{2}$ bins 
are used for the other studied samples.  
The parameterizations and $\Delta E\!- \!M_{bc}$ 
ranges considered in these fits
are the same as those used for the total $B$ yield extraction. 
The widths and means of the Gaussians describing
the signal are fixed at the values obtained for the total signal sample, while
the signal yield and the background PDF's parameters are free parameters.

The background-subtracted $M(D^{0} \bar{D}^{0})$ in the
$\psi(3770)$ signal region is shown in 
Fig.~\ref{d0d0k_mass_bins}(a).
The peak is fitted for $M(D^{0} \bar{D}^{0}) < 4~\rm{GeV}/{\it c}^{2}$ 
with a Breit-Wigner (BW) plus a threshold function to describe a
phase space component. The $\psi(3770)$ signal yield is $68 \pm 15$
events with a peak mass of $3776\pm 5~\rm{MeV}/{\it c}^{2}$, 
and a width of $27 \pm 10~\rm{MeV}/{\it c}^{2}$,
in agreement with the PDG averages.
\begin{figure}[!hb]      
\vspace{-0.1cm}
\begin{center}      
\includegraphics[width=0.23\textwidth,height=0.19\textwidth]{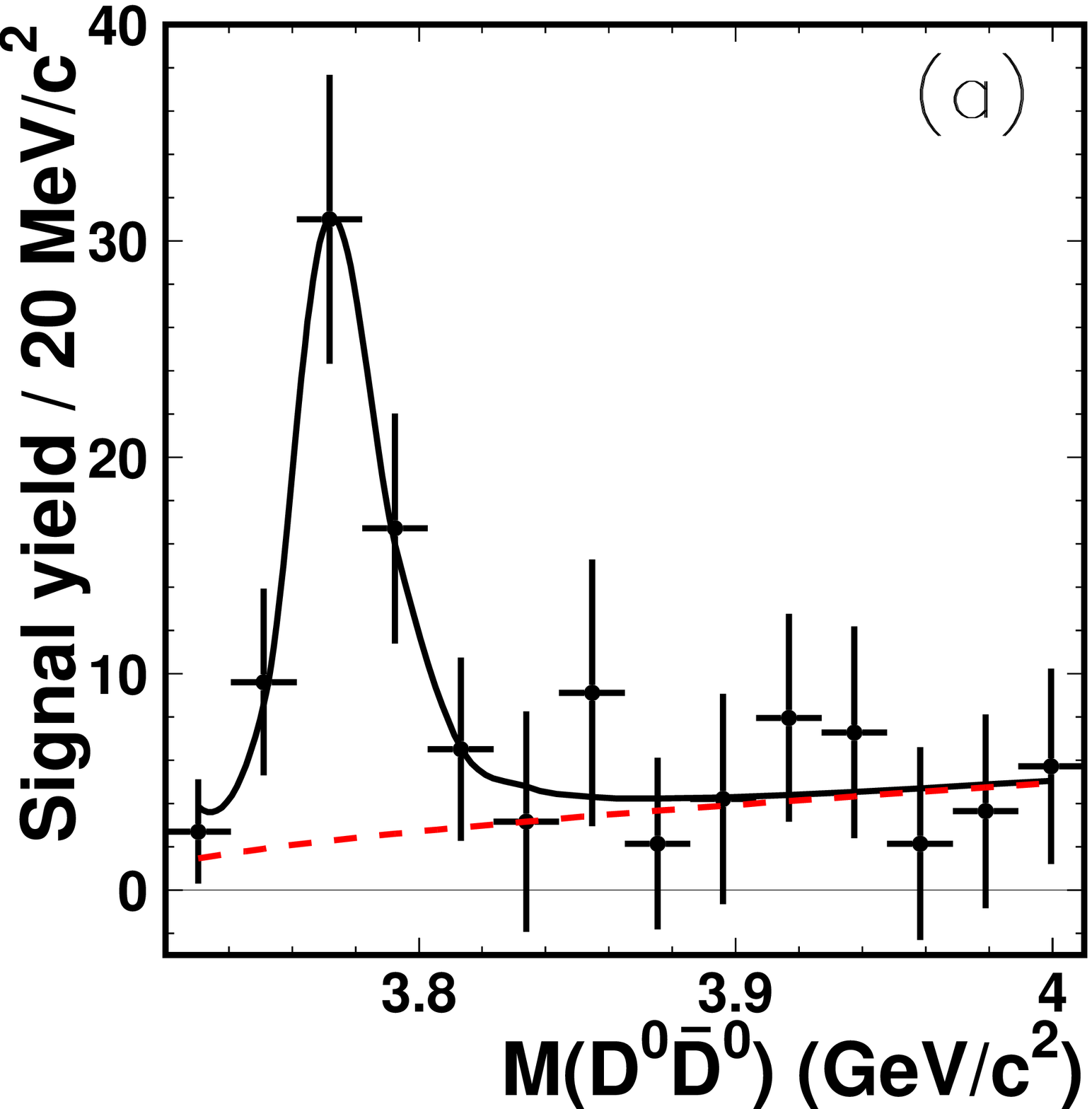}   
\includegraphics[width=0.23\textwidth,height=0.19\textwidth]{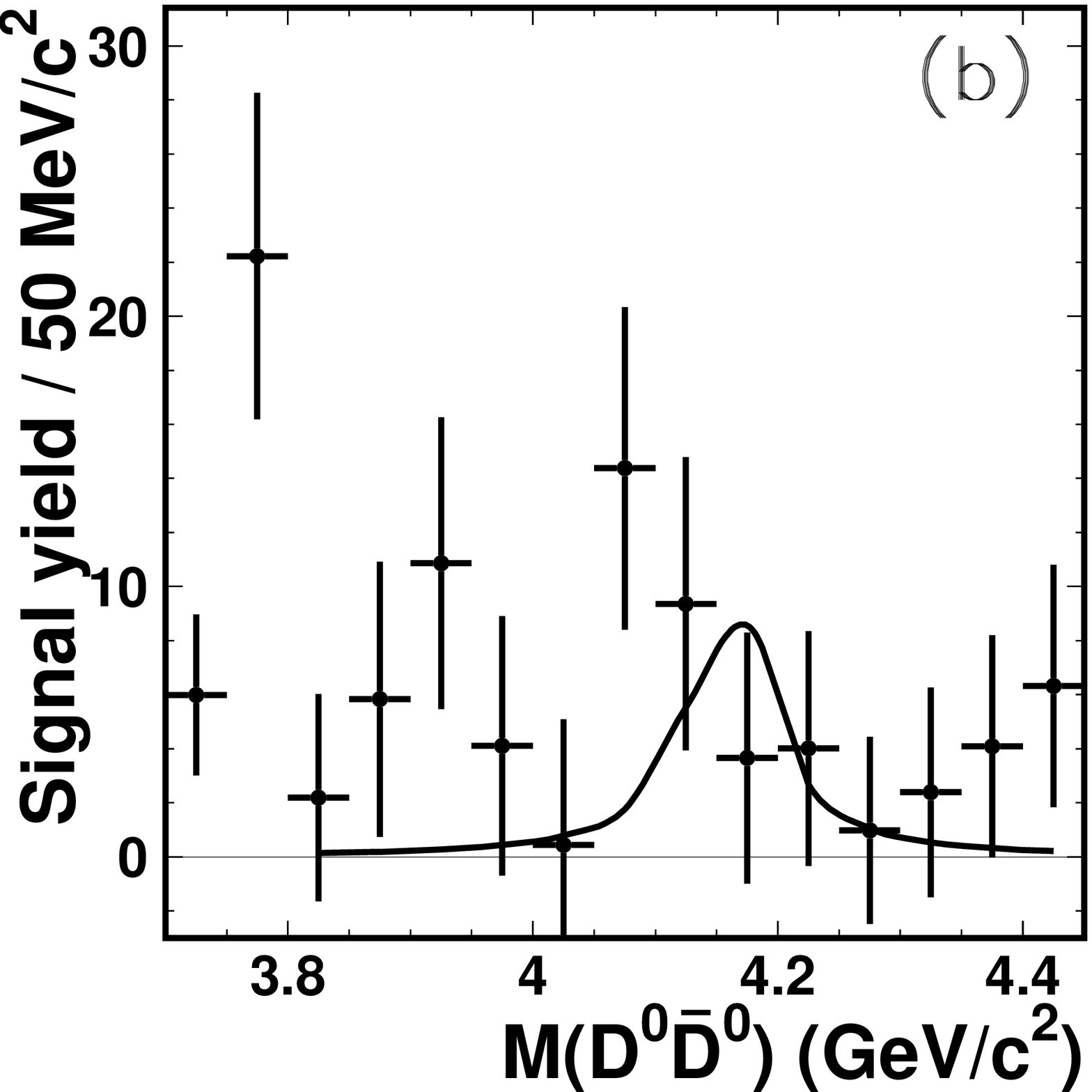}
\includegraphics[width=0.23\textwidth,height=0.19\textwidth]{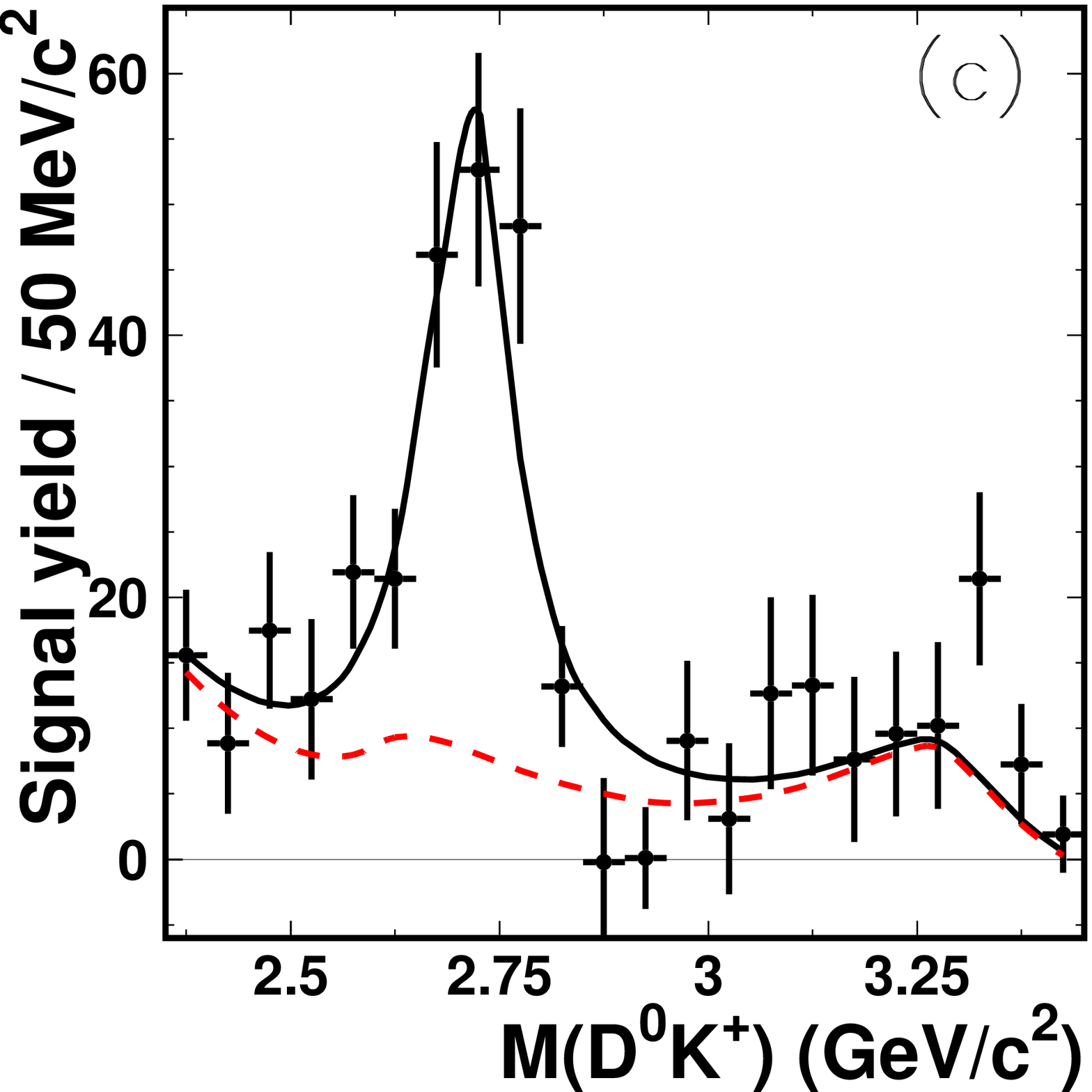}    
\caption{
$B$ meson signal yield vs.: 
         (a) $M(D^{0} \bar{D}^{0})$ in the $\psi(3770)$ region, 
         (b) $M(D^{0} \bar{D}^{0})$ for $\cos \theta_{\rm{hel}}>0$
         (c) $M({D}^{0}K^{+})$ for $M(D^{0}\bar{D}^{0}) > 3.85~\rm{GeV}/{\it c}^{2}$. 
Solid curves denote the $\chi ^2$ fit results described in the text. 
The red/dotted curve in (a) shows the phase-space component, 
whereas in (c) the red/dotted curve is the sum of the three components: 
$\psi(4160)$ reflection, phase-space and threshold components. }
\label{d0d0k_mass_bins}
\end{center}      
\vspace{-0.6cm}
\end{figure}

The background-subtracted $M(D^{0}\bar{D}^{0})$ spectrum 
(Fig.~\ref{d0d0k_mass_bins}(b)), for events satisfying 
$\cos \theta_{\rm{hel}} > 0$, where $\theta_{\rm{hel}}$ is the helicity angle 
between the $D^{0}$ momentum vector 
and the direction opposite the $K^+$ in the $D^{0} \bar{D}^{0}$  
rest frame, is used to estimate the possible $\psi(4160)$, 
$\psi(4040)$ contribution to the enhancement 
at $M(D^{0}K^{+}) \! \simeq \! 2.7~\rm{GeV}/{\it c}^2$. 
The peak at threshold corresponds to $\psi(3770)$, while the 
structure 
at $4.0 \div 4.2 ~\rm{GeV}/{\it c}^2$ is conservatively 
assummed to be predominantly (\cite{bes_cb},~\cite{barnes}) 
due to the $\psi(4160)$. 
The distribution for $M(D^{0}\bar{D}^{0}) > 3.8~\rm{GeV}/{\it c}^{2}$
is fitted with a BW with mass and width 
fixed at the nominal $\psi(4160)$ values 
($M=4160$, $\Gamma = 80~\rm{MeV}/{\it c}^{2}$~\cite{pdg04}),  
yielding $24 \pm 11$
signal events. We use these $\psi(4160)$ parameters 
to estimate the number of $\psi(4160)$ events in the backward 
helicity-angle hemisphere, in the region
$M({D}^{0}K^{+}) < 2.9~\rm{GeV}/{\it c}^{2}$.
Taking into account the efficiency we obtain a total of 
$43 \pm 20$ $\psi(4160)$ events.

Figure~\ref{d0d0k_mass_bins}(c) shows the
background-subtracted 
$M({D}^{0}K^{+})$
distribution for events with $M(D^{0}\bar{D}^{0}) > 3.85~\rm{GeV}/{\it c}^{2}$.
This requirement removes the $\psi(3770)$ reflection at 
high $M({D}^{0}K^{+})$. The predicted $\psi(4160)$ reflection agrees
well with the data in the high mass $M({D}^{0}K^{+})$ region but 
does not explain the large peak at $M({D}^{0}K^{+})\simeq 2.7~\rm{GeV}/{\it c}^{2}$.
We parameterize the observed excess of events with a BW
and fit the $M({D}^{0}K^{+})$ spectrum (Fig.~\ref{d0d0k_mass_bins}(c))
with the ansatz of a new resonance,
the $\psi(4160)$ reflection and a 
phase-space component with shapes determined by MC simulations.
The efficiency variation is taken into account in the fit; 
the free parameters are the resonance yield, mass and width, 
and the phase-space component normalization.
The fit has an acceptable overall $\chi ^2$ but is unable to reproduce
the events near the low-mass threshold seen in Fig.~\ref{d0d0k_mass_bins}(c).
We used several phenomenological 
parameterizations
(polynomials, a BW, an exponential) of the threshold enhancement
in the fit to determine its influence on the BW parameters 
of the $2.7~\rm{GeV}/{\it c}^{2}$ peak.
The exponential form $a\times \exp {[-\alpha M^2(D^{0}K^{+})]}$ 
gives a good description of the mass spectrum, while adding only two
free parameters. 
 
For the new resonance, which we henceforth denote 
as the  $D_{sJ}(2700)^{+}$, we obtain 
a signal yield of $182 \pm 30$ events, a mass of 
$M = 2708 \pm 9~\rm{MeV}/{\it c}^{2}$ and a width of  
$\Gamma = 108 \pm 23~\rm{MeV}/{\it c}^{2}$. 
The threshold and the phase-space components from the
fit are $58 \pm 38$ and $47\pm 26$ events, respectively.
The fit results are shown in 
Figs.~\ref{xx_d0d0bkc_plus_reflections}(a)-(c)
as histograms overlaid on the measured mass spectra. 

The resonance parameters and product branching fractions 
are summarized in Table~\ref{bf_table} (the first error is statistical, 
the second is systematic). 
The systematic errors on the product branching fractions 
and the resonance parameters
include contributions from uncertainties in the yields of 
the $\psi(4160)$ reflection 
(including the recent $\psi(4160)$ parameter determination~\cite{bes_cb}), 
the threshold parameterization,  
sensitivities of parameters to the fit range and 
parameterization, uncertainties in the ${\cal LR}_B$ selection, 
as well as uncertainties due to interference effects that were neglected. 
The systematics due to the latter  are
determined from MC simulations of Dalitz plot densities
with and without interference of contributing amplitudes, 
with each contributing resonance parameterized by a BW form. 
The resonance parameters from Table~\ref{bf_table}
and the threshold enhancement parameters are used to 
determine the amplitudes.
The effects of interference of the $\psi(3770)$ with other 
states are found to be small
and are neglected in the simulations.
These MC samples, with maximal constructive and destructive interferences, 
were analysed ignoring interference effects.
The differences between the obtained resonance parameters
and the input values are taken as systematic errors. \\
\begin{table}[]
\vspace{-0.1cm}
\caption[] {Resonance parameters and product branching fractions:
${\cal B}(B^+ \! \to \! \bar{D}^0 D_{sJ}(2700)^{+}) 
\times  {\cal B}(D_{sJ}(2700)^{+} \! \to \! D^0 K^+)$ and 
${\cal B}(B^+ \! \to \! \psi(3770) K^+ ) 
\times  {\cal B}(\psi(3770)  \! \to \! D^0 \bar{D}^0)$.
}
\begin{tabular}{l| c| c } 
$    R         $          & $D_{sJ}(2700)^{+}$ 
                         & $\psi(3770)$
                                                 \\ \hline \hline
${\rm N_{{\rm sig}}}$ (${\rm Significance}$) \       
                         &\ $ 182 \pm 30$  \ ($8.4 \sigma $) \
                         & \ $ 68 \pm 15$   \ ($5.5 \sigma $) \
                                                 \\ 
${\rm M \ [MeV/{\it c}^{2}]}$      & $ 2708 \pm 9^{+11}_{-10}$
                         & $ 3776 \pm 5  \pm 4$ 
                                                 \\ 
${\rm \Gamma \ [MeV/{\it c}^{2}]}$& $ 108 \pm 23^{+36}_{-31}$
                         & $ 27 \pm 10 \pm 5$ 
                                                 \\ \hline
Product ${\cal B}$  $[10^{-4}]$   
                         & $ 11.3 \pm 2.2^{+1.4}_{-2.8} $  
                         & $ 2.2 \pm 0.5 \pm 0.3$  \\
\end{tabular}
\label{bf_table}
\vspace{-0.2cm}
\end{table}
\begin{figure}[]      
\vspace{-0.1cm}
\begin{center}      
\includegraphics[width=0.23\textwidth,height=0.19\textwidth]{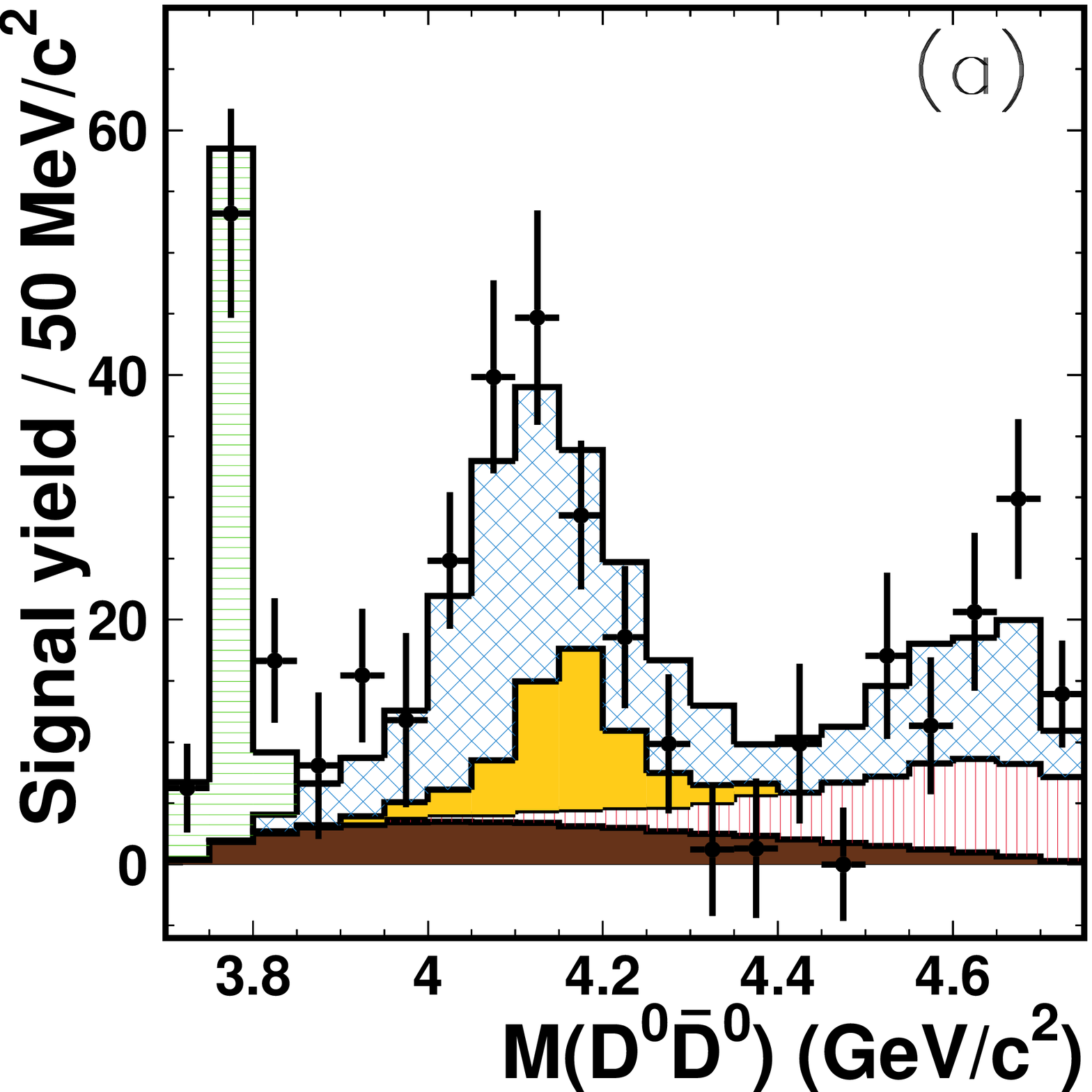}      
\includegraphics[width=0.23\textwidth,height=0.19\textwidth]{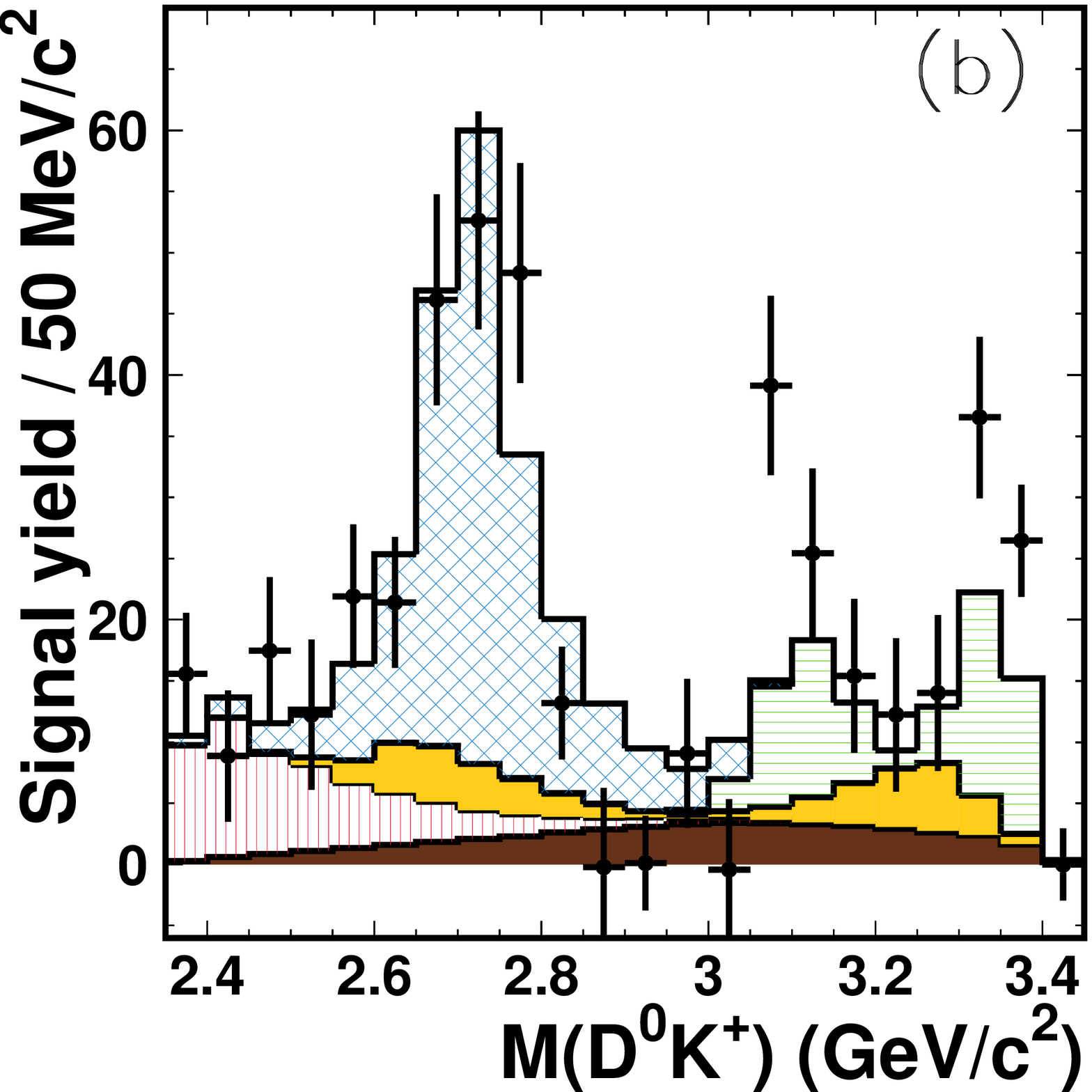}      
\includegraphics[width=0.23\textwidth,height=0.19\textwidth]{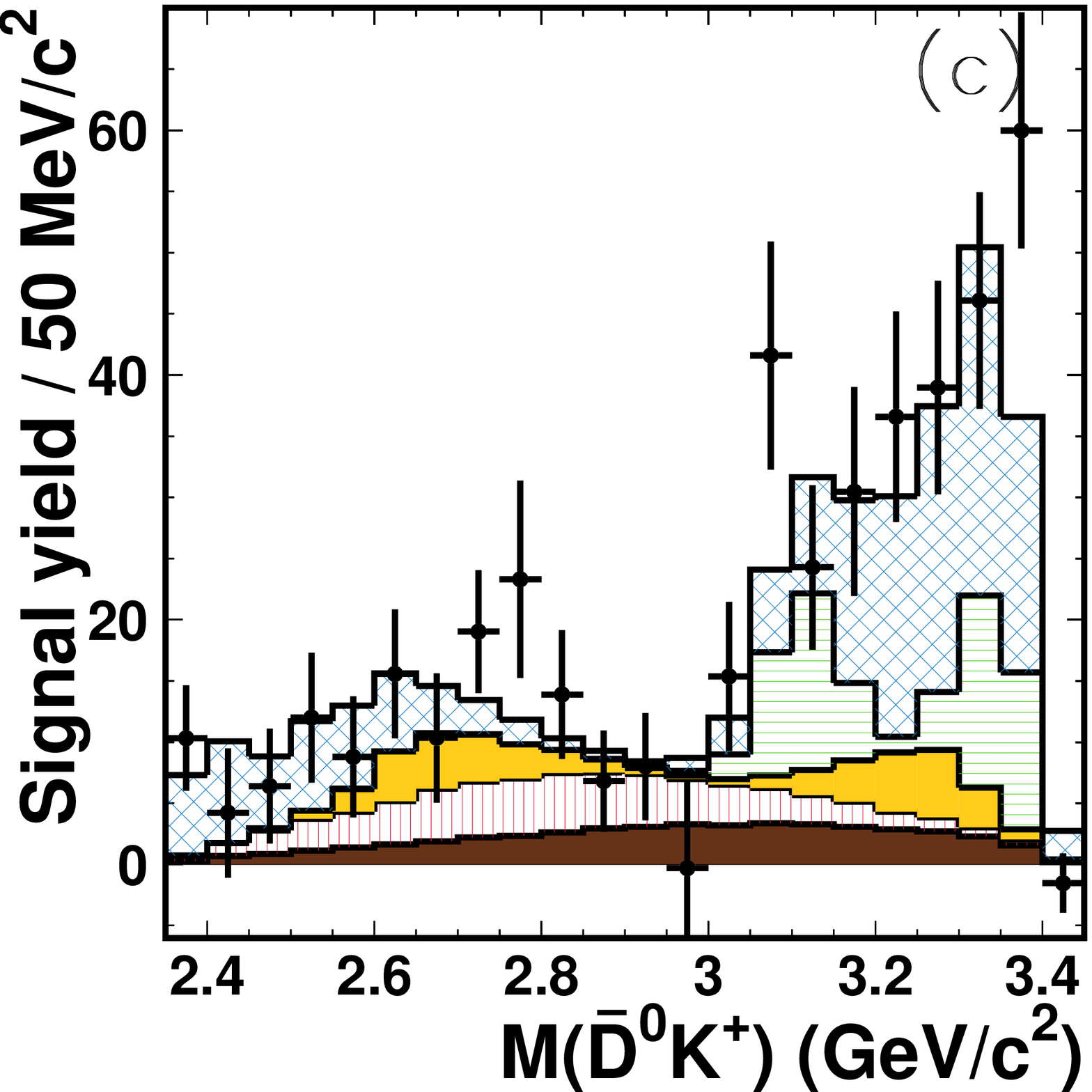}     
\includegraphics[width=0.23\textwidth,height=0.19\textwidth]{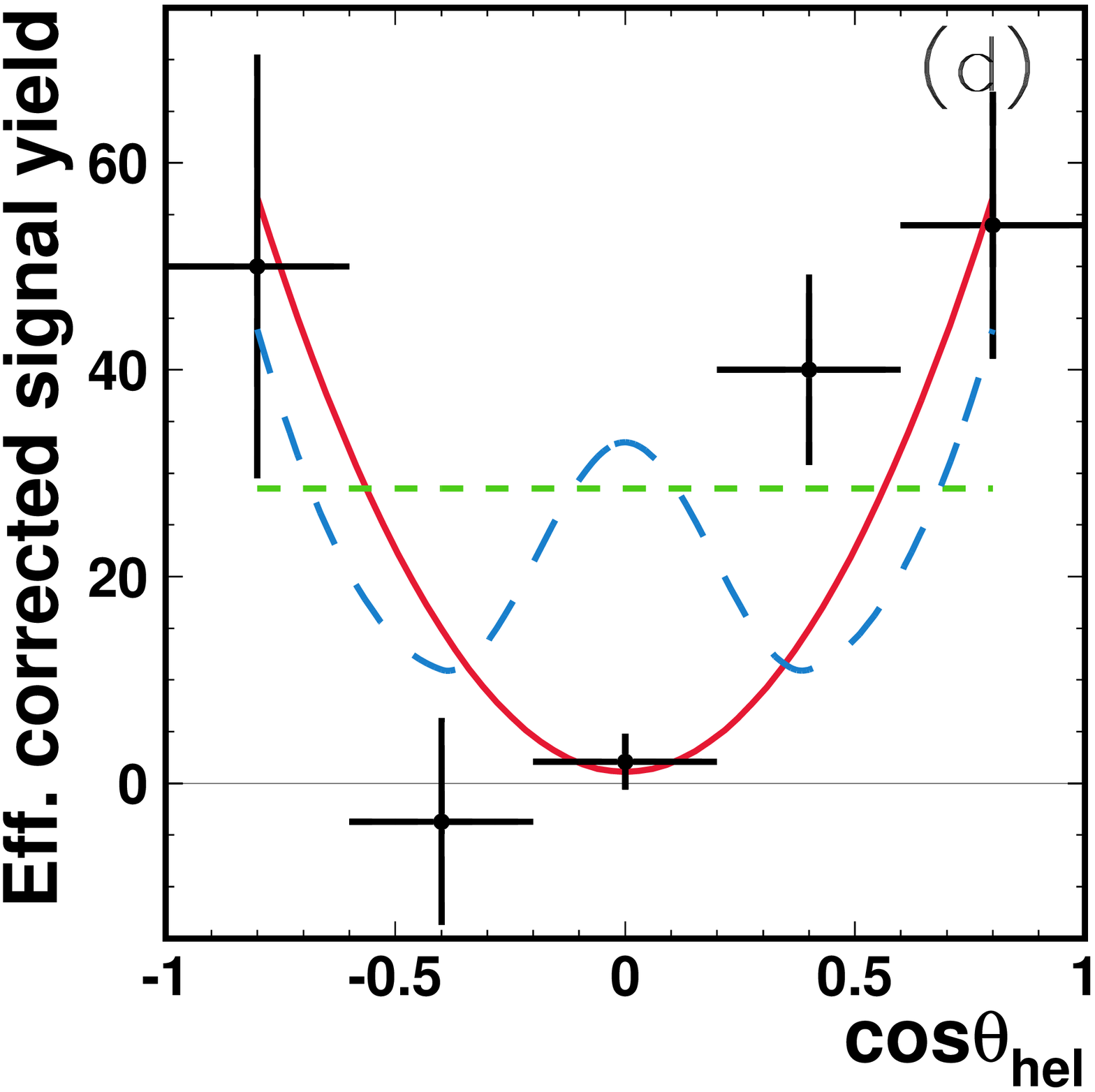}
\caption{$B$ meson signal yield vs. :
         $M(D^{0}\bar{D}^{0})$~(a), $M({D}^{0}K^{+})$~(b),
         and $M(\bar{D}^{0}K^{+})$~(c).
         Histograms denote the contributions from:
         $D_{sJ}(2700)^{+}$ (blue/grid),
         $\psi(3770)$ (green/horizontally striped), 
         $\psi(4160)$ (yellow/light grey), 
         threshold (red/vertically striped) and  
         phase-space components (brown/dark grey).
         Histograms are superimposed additively. 
         (d) Efficiency corrected $D_{sJ}(2700)^{+}$ helicity angle 
         distribution. Curves show  
         predictions for various spin hypotheses: 
         $J \! = \! 0$ (green/dotted line), 
         $J \! = \! 1$ (red/solid), $J \! = \! 2$ (blue/dashed), 
         where Legendre polynomial values are averaged over the bin width.} 
\label{xx_d0d0bkc_plus_reflections}
\end{center}      
\vspace{-0.6cm}
\end{figure}      
We study background-subtracted $\psi(3770)$  and 
$D_{sJ}(2700)^{+}$ helicity angle distributions
by selecting the respective invariant mass in the resonance
region and obtaining $B$ meson 
signal yields in bins of $\cos \theta_{\rm{hel}}$
from the 
2D fits to $\Delta E$ and $M_{\rm bc}$. 
Here $\cos \theta_{\rm{hel}}$ for $\psi(3770)$ 
is defined as before, 
whereas for $D_{sJ}(2700)^{+}$ it is the angle between 
the $K^+$ momentum vector and the direction opposite 
the $\bar{D}^{0}$ in the $D^{0}K^+$ rest frame. 
The obtained angular distributions are then corrected using 
bin-by-bin efficiencies. 
The expected reflections from $\psi(4160)$ and from the threshold 
component are subtracted from the $D_{sJ}(2700)^{+}$ angular 
distribution.  
Spin hypotheses for the resonances are tested by 
comparing predictions for the different hypotheses   
to the corrected angular distributions.
The $\psi(3770)$ distribution (not shown) is well described   
by the $J=1$ hypothesis ($\chi^{2}/ndf=3.6/5$).
The $D_{sJ}(2700)^{+}$ distribution (Fig.~\ref{xx_d0d0bkc_plus_reflections}(d))
favours $J=1$ ($11/5$); the $J=0$ ($112/5$) and $J=2$ ($146/5$) 
assignments can be rejected. 
The $J=1$ assignment and the observed decay to two pseudoscalar mesons
imply parity $P=-1$. 

In summary, from a study of the Dalitz plot we find that the decay  
$B^{+} \! \to  \! \bar{D}^{0}D^{0}K^{+}$ proceeds dominantly via
quasi-two-body channels: 
$B^{+} \! \to \! \bar{D}^{0} D_{sJ}(2700)^{+}$ and
$B^{+} \! \to \! \psi(3770) K^{+}$. 
The observed rate for $\psi(3770)$ production in $B$ meson decays
confirms our previous observation~\cite{chistov}.
The $D_{sJ}(2700)^{+}$ is a previously unobserved
resonance in the $D^{0}K^{+}$ system with a mass 
$M=2708 \pm 9 ^{+11}_{-10}~\rm{MeV}/{\it c}^{2}$, 
width  $\Gamma = 108 \pm 23 ^{+36}_{-31} ~\rm{MeV}/{\it c}^{2}$
and $J^P=1^-$.  
The statistical significance of this observation is $8.4 \sigma$.  
Based on its observed decay channel,
we interpret the  $D_{sJ}(2700)^{+}$
resonance as a $c\bar{s}$ meson.
Potential model calculations~\cite{godfrey-isgur} predict a $c\bar{s}$
radially excited  $2^{3}S_{1}$ state with a mass $2710$-$2720~\rm{MeV}/{\it c}^{2}$.
From chiral symmetry considerations~\cite{maciek} a $1^{+}$-$1^{-}$ 
doublet of states  has been predicted.  If the $1^+$ state
is identified as the $D_{s1}(2536)$, the mass predicted for the $1^-$ state
is $M= 2721\pm 10~\rm{MeV}/{\it c}^{2}$. 
Additional measurements of the meson properties are needed to distinguish between
these two interpretations. 

It is not clear whether the structure at $2688~\rm{MeV}/{\it c}^{2}$ observed 
recently~\cite{babar_ds2860} 
in the $DK$ system produced in continuum could be due to the $D_{sJ}(2700)^{+}$. 
The recently reported $D_{sJ}(2860)$ state~\cite{babar_ds2860} is not seen in our data.
This could indicate a high spin for this meson that suppresses its production in
$B$ decays.

We thank the KEKB group for excellent operation of the
accelerator, the KEK cryogenics group for efficient solenoid
operations, and the KEK computer group and
the NII for valuable computing and Super-SINET network
support.  We acknowledge support from MEXT and JSPS (Japan);
ARC and DEST (Australia); NSFC and KIP of CAS (China); 
DST (India); MOEHRD, KOSEF and KRF (Korea); 
KBN (Poland); MES and RFAAE (Russia); ARRS (Slovenia); SNSF (Switzerland); 
NSC and MOE (Taiwan); and DOE (USA).



\begin{thebibliography}{99}
\bibitem{cc}
Throughout this paper, 
the inclusion of the charge conjugate mode decay is implied.

\bibitem{aleph}
R.~Barate {\it et al.} (ALEPH Collaboration), 
Eur. Phys. J. C {\bf 4}, 387 (1998).
 
\bibitem{babar}
B.~Aubert {\it et al.} (BaBar Collaboration),
Phys. Rev. D {\bf 68}, 092001 (2003).

\bibitem{KEKB}
S.~Kurokawa and E.~Kikutani, Nucl. Instr. and Meth. A {\bf 499}, 1 (2003),
and other papers included in this volume.

\bibitem{Belle}
A.~Abashian {\it et al.} (Belle Collaboration),
Nucl. Instr. and Meth. A {\bf 479}, 117 (2002).

\bibitem{fox-wolfram}
G.~C.~Fox and S.~Wolfram, Phys. Rev. Lett. {\bf 41}, 1581 (1978).

\bibitem{argus}
H.~Albrecht {\it et al.} (ARGUS Collaboration),
Phys. Lett. B {\bf 229}, 304 (1989).

\bibitem{pdg04}
S.~Eidelman {\it et al.}, Phys. Lett. B {\bf 592}, 1 (2004).

\bibitem{bes_cb}
W.-M. Yao {\it et al.}, J. Phys. G {\bf 33}, 1 (2006);
M.~Ablikim {\it et al.} (BES Collaboration), arXiv:0705.4500v1 [hep-ex].

\bibitem{barnes}
T.~Barnes, S.~Godfrey, E.~S.~Swanson, Phys. Rev. D {\bf 72}, 054026 (2005).

\bibitem{chistov}
R.~Chistov {\it et al.} (Belle Collaboration), 
Phys. Rev. Lett. {\bf 93}, 051803 (2004).

\bibitem{godfrey-isgur}
S.~Godfrey and N.~Isgur,
Phys. Rev. D {\bf 32}, 189 (1985);
F.~E.~Close {\it et al.}, Phys. Lett. B {\bf 647}, 159 (2007).

\bibitem{maciek}
M.~A.~Nowak, M.~Rho and I.~Zahed,
Acta Phys. Polon. B {\bf 35}, 2377 (2004).

\bibitem{babar_ds2860}
B.~Aubert {\it et al.} (BaBar Collaboration),
Phys. Rev. Lett. {\bf 97}, 222001 (2006).

\end{thebibliography}
\end{document}